\documentclass[11pt]{article}
\usepackage{amsmath}
\usepackage{setspace}
\usepackage[psamsfonts]{amsfonts}
\usepackage{theorem}
\usepackage{version}
\usepackage{graphicx}
\usepackage{pstricks,pst-node,pst-text,pst-3d}
\usepackage{subfigure}

\newtheorem{lemma}{Lemma}[section]

\newtheorem{theorem}{Theorem}[section]

\newcommand{\cqfd}{\hfill $\square$}

\newcommand{\N}{\mathbb N}

\newcommand{\pla}{\ensuremath{A}}
\newcommand{\plb}{\ensuremath{B}}
\newcommand{\plc}{\ensuremath{C}}

\newcommand{\Nn}{\mathbb{N}}
\newcommand{\al}{\alpha}
\newcommand{\be}{\beta}

\newcommand{\twenty}{\textwidth 16.2cm}

\begin{document}
\title {{\sc A Stochastic Analysis of Table Tennis}}

\author{Yves {\sc Dominicy}\thanks{Research supported by an IAP P6/07 contract, from the IAP program (Belgian Scientific Policy): Economic policy and finance in the global economy.} 
, \
Christophe {\sc Ley}\thanks{Research supported by a Mandat 
d'Aspirant of the Fonds National de la Recherche Scientifique, 
Communaut\' e fran\c caise de Belgique.} 
\ \ and \ 
Yvik {\sc Swan}\thanks{Research supported by a Mandat 
de Charg\'{e} de Recherches of the Fonds National de la Recherche Scientifique, 
Communaut\' e fran\c caise de Belgique.} 
\\
{\it \small
E.C.A.R.E.S. and D\' epartement de Math\' ematique
}
\\
{\it \small
Universit\' e Libre de Bruxelles, \twenty 
Brussels, Belgium
}
\\
$\;$\vspace{-1mm}\\
}
\date{}

\maketitle

\vspace{-10mm}
\begin{abstract}\twenty
We establish a general formula for the distribution of the score in table tennis.  We use this formula to derive the probability distribution (and hence the expectation and variance) of the number of rallies necessary to achieve any given score. We use these findings to investigate  the dependence of these quantities on the different parameters involved (number of points needed to win a set, number of consecutive serves, etc.), with particular focus on the rule change imposed in 2001 by the International Table Tennis Federation (ITTF). Finally we briefly indicate how our results can lead to more efficient estimation techniques of individual players' abilities.
\end{abstract}\twenty

\section{Introduction.}\label{intro}

We consider the following situation. Two players (or teams), referred to as $\pla$ and $\plb$, play a sequence of \emph{rallies} after each of which either $\pla$ or $\plb$ is declared winner. Every rally is initiated by a \emph{server}---the other player  is then called the \emph{receiver}---and  a point is scored (by the winner) after each rally. The players continue competing until a \emph{match}-winner is declared, with a match being composed of several \emph{sets} according to the $(m,n, G)$-scoring system, which we define as follows.
 
\begin{quote} \emph{The $(m, n, G)$-scoring system}:   (i) A \emph{set}  consists of a sequence of independent  rallies; the winner of a set is the first player  to score $n$ points or, in case of a \emph{tie} at  $n-1$,  the first player  to create a difference of two points after the tie.  (ii)  A \emph{match}  consists of a sequence of  independent  {sets};  the winner of a match is the first player  to win $G$ sets. (iii) The server in the first rally of the first set is determined by flipping a coin;  if $G\ge2$, the right to serve in the first rally of each subsequent set alternates between the two opponents.  (iv) Within a set, the right to serve changes between the opponents after each sequence of $m$ consecutive rallies by a server until either a set-winner is declared or a tie is reached at $(n-1, n-1)$. (v) After a tie at $(n-1, n-1)$, the right to serve alternates after each rally. \end{quote}

To the best of our knowledge, table tennis is the only sport currently using the  $(m,n,G)$-scoring system, although speed badminton (a.k.a. speedminton or, before 2001, shuttleball)  can nevertheless be shown to fit, up to a minor modification, within the above scoring system for $m=3$, $n=16$ and $G=2$ (the change concerns the right to serve in the first rally of each subsequent set, as here the player who has lost the previous set serves first). 
Until 2001, table tennis was played according to the $(5, 21, G)$-scoring system, with $G = 2$ or 3. In 2001 the  International Table Tennis Federation (ITTF) decided to switch to the $(2,11,G)$-scoring system, with $G$ either 3 or 4. 

In this article we regard table tennis as a succession of identical and independent random experiments (the rallies) and analyze the properties of the  random processes (score change, game duration, etc.) induced by the scoring system. 
%
%
As in most of the (mathematical) literature on this kind of sport  we will restrict our attention to two well-known models. The first is the so-called \emph{server model} in which it is assumed that rally-outcomes are mutually independent and are, conditionally upon the server, identically distributed. Denoting by $p_a$ (resp., $p_b$) the probability that player~$\pla$ (resp., player~$\plb$) wins a rally he/she initiates, the game is then entirely governed by the bivariate parameter $(p_a,p_b)\in(0,1)\times(0,1)$. The second model under investigation is the simpler  \emph{no-server model}, in which it is assumed that rallies are won   with probabilities that do not depend on the server; the latter model is thus a particular case of the former, with parameters  $p_a=1-p_b=p\in(0,1)$. For a discussion on the validity of  these  models we refer the reader to  \mbox{e.g.} Klaassen and Magnus (2001) where is is shown, within the framework of the game of tennis,  that i.i.d. models provide a good approximation to real games.  We have not investigated whether their conclusions can be transposed to our framework, and reserve such empirical considerations for later publications.

While an important literature has been devoted to  sports such as  tennis (see, e.g., Hsi and Burich~1971,  George~1973, Carter and Crews~1974, Riddle~1988 or Klaassen and Magnus~2003) or badminton  (see, e.g., Phillips~1978, Simmons~1989 or the recent contributions by Percy~2009 and Paindaveine and Swan~2011), the corresponding literature on table tennis is, to say the least, scant. Indeed,  to the best of our knowledge, only one paper (Schulman and Hamdan~1977) addresses this issue and proposes a solution only for  $(m,n,G)=(5,21,2)$. 
%
This  lack of interest  perhaps originates in the apparent simplicity of the stochastics underlying the game of table tennis. Indeed,  whereas in sports such as badminton or tennis the number of consecutive serves by any player  is random, the right to serve  in table tennis changes  according to a deterministic rule, and thus rather simple (albeit delicate) combinatoric arguments allow to obtain all the corresponding probability distributions -- see Section~\ref{lemmasect}. Moreover the formulae we obtain are -- see  Appendix A.1 -- rather lengthy and cumbersome. 

This dismissive view of the problem is, in our opinion, wrong. Consequently, we tackle in this paper all possible questions of interest related to sports based on the $(m, n, G)$-scoring system, and solve them in full generality. The  formulae we obtain are based on rally-level combinatorial arguments and allow us to derive the probability distribution (and hence the expectation and variance) of the number of rallies necessary to achieve \emph{any} given succession of rally-outcomes. By doing so, we extend the previous contribution on the subject (Schulman and Hamdan 1977) where only  the special case of the $(5,21,2)$-scoring system and only set-winning probabilities are computed.  

The outline of the paper is as follows. In Section~\ref{lemmasect}  we fix the notations, obtain the main theoretical results and derive the distributions of the score and game-duration.  In Section~\ref{compare} we    investigate the dependence of the quantities obtained in Section~\ref{lemmasect} on the parameters $(m, n, G)$, with a particular attention given to the effects of the ITTF's rule change (see above) on the game-winning probabilities and durations. Some final comments and an outlook on future research are stated in Section~\ref{fc},  while an appendix collects the full-length formulae and their  proofs.

\section{Distribution of the score,  set-winning probabilities and distribution of the number of rallies in a single set.}\label{lemmasect}


In this section we fix the notations and obtain the fundamental probabilities associated with  the scoring process within a single set (see Lemma \ref{terrible} below). We use these  to derive the distribution of the scores, the set-winning probabilities as well as the distribution of the number of rallies needed to complete a set. 
\subsection{Notations} 
Here and  throughout we denote scores by couples of integers where the first entry (resp., the second entry) stands for the number of points scored by player~$\pla$ (resp., by player~$\plb$). We will reserve the use of the notation $(n,k)$ (resp., $(k,n))$ to indicate the final score in a set won by $\pla$ (resp., by $\plb$) without a tie. General intermediate scores will be denoted $(\alpha, \beta)$. 

We call $\pla$-set (resp., $\plb$-set) a set in which player~$\pla$ (resp., player  $\plb$) is the first server. Note the symmetric roles played by $\pla$ and $\plb$, which will often allow us to state our results in terms of $\pla$-sets only. For $\plc_1, \plc_2 \in \{\pla, \plb\}$, we denote by $p_{\plc_1}^{\plc_2}$ the probability that $\plc_2$ wins a $\plc_1$-set. Now, similarly to Schulman and Hamdan~(1977), it can be shown that, for $m$ dividing $n-1$, $p_\pla^\pla=p_\plb^\pla$ (resp., $p_\pla^\plb=p_\plb^\plb$), \black i.e. the probability for each player to win a single set is not affected by the choice of the first server in the set. Consequently, the probability of player~$\pla$ (resp., player~$\plb$) winning a match can be obtained in a straightforward manner by  simple conditioning. Since the values of $m$ chosen by international federations always satisfy the aforementioned  divisibility constraint, we can (and will) restrict most of our attention on the outcome of a single $\pla$-set. 

For $\plc\in\{\pla, \plb\}$, we denote by $E_\pla^{\al,\be,\plc}$ the event associated with all sequences of $\alpha+\beta$ rallies  in an $\pla$-set that gives rise to $\al$ (resp., $\be$) points scored by player~$\pla$ (resp., by player~$\plb$) and   sees player~$\plc$ score the last point. The latter condition   entails that we necessarily assume $\al>0$ (resp., $\be>0$) when $\plc=\pla$ (resp., when $\plc=\plb$). 
We denote the corresponding probability by 
\begin{equation}\label{probaE}
p_{\pla}^{\al,\be,\plc}={\rm P}[E_\pla^{\al,\be,\plc}]. 
\end{equation}

With these notations, an $\pla$-set is won by $\pla$ (resp., by $\plb$) on the score $(n, k)$ (resp., $(k,n)$) with probability $p_{\pla}^{n,k,\pla}$ (resp., $p_{\pla}^{k,n,\plb}$), for $k\le n-2$.   Obviously these quantities do not suffice to compute the probability $p_\pla^\plc$ that $\plc$ wins a set initiated by $\pla$, since we still need to account for what happens in case of a tie at $n-1$, an event which occurs with probability $p_\pla^{n-1,n-1,\pla}+p_\pla^{n-1,n-1,\plb}$. For this we  introduce the notation     $p_\pla^{{\rm tie},\pla}$ (resp.,  $p_\pla^{{\rm tie},\plb}$) to represent the probability that $\pla$ (resp., $\plb$) scores 2 more  consecutive points than his/her opponent after the tie; this quantity will have to be computed differently than those defined above,  since the rules governing the game after a tie are not the same as those which were applicable before this event. With these notations we get 
\begin{equation} \label{pAA} p_\pla^\pla=\sum_{k=0}^{n-2}p_{\pla}^{n,k,\pla}+(p_{\pla}^{n-1,n-1,\pla}+p_{\pla}^{n-1,n-1,\plb})p_\pla^{{\rm tie},\pla},\end{equation}
and  
\begin{equation} \label{pAB} p_\pla^\plb=\sum_{k=0}^{n-2}p_{\pla}^{k,n,\plb}+(p_{\pla}^{n-1,n-1,\pla}+p_{\pla}^{n-1,n-1,\plb})p_\pla^{{\rm tie},\plb}.\end{equation}
The corresponding quantities for $\plb$-sets are defined by switching the roles of the two players in these definitions.

Determining (\ref{probaE}) and computing $p_\pla^{{\rm tie},\plc}$ is the key to our understanding of  the problem. We derive a general formula for these quantities in the following section. 

\subsection{Distribution of the scores in a single set.}\label{fund_res}
In order to give the reader a feeling for what kind of mechanics are at play, we start with the  (very simple)     no-server model. Here every rally is won by player  $\pla$ (resp., by player  $\plb$) with probability $p$ (resp., $1-p$), irrespective of the server. The process therefore reduces to a succession of independent Bernoulli trials with success-probability $p$, yielding
%
%
\begin{equation}\label{super}p_\pla^{\alpha,\beta,\pla}=\displaystyle{\binom{\al+\be-1}{\al-1} p^{\al}(1-p)^{\be}} \mbox{ and }p_\pla^{\alpha,\beta,\plb}=\displaystyle{\binom{\al+\be-1}{\al} p^{\al}(1-p)^{\be}}\end{equation}
  for all $\alpha, \beta \in \N$. 
%
All that remains is to  determine the probabilities $p_A^{{\rm tie},A}$ and $p_A^{{\rm tie},B}$. Clearly, the number of rallies played after a tie until either player wins the set has to be even. Let $\ell\in\N_0$. In order to have $2\ell$ rallies after the tie, it suffices that $\pla$ and $\plb$ win alternatively during the first  $2(\ell-1)$ rallies, and that one of the two players scores 2 successive points thereafter. This entails that the probability that $\pla$ (resp., $\plb$) wins the set after $2\ell$ points is given by $p^2(2p(1-p))^{\ell-1}$ (resp., by $(1-p)^2(2p(1-p))^{\ell-1}$). \black Hence the probability of $\pla$ (resp., of $\plb$)  winning the set after a tie at $n-1$ is given by 
\begin{equation}\label{super2}p_\pla^{{\rm tie},\pla}=\frac{p^{2}}{1-2p(1-p)} \mbox{ and } p_\pla^{{\rm tie},\plb}=\frac{(1-p)^{2}}{1-2p(1-p)}.\end{equation}
Thus,   from (\ref{pAA}), (\ref{pAB}) and  \eqref{super}, simple summation yields the game-winning probabilities $p_\pla^\pla$ and $p_\pla^\plb$.  Note that these quantities do not depend on the parameter $m$. 

The case of  the server model is trickier. The scoring process of $\pla$  (resp., of $\plb$) is   a succession of $m$ independent Bernoulli trials with success-probability $p_a$ (resp., $1-p_a$), followed by  $m$ independent trials with success-probability $1-p_b$ (resp., $p_b$), followed again by $m$ Bernoulli trials identical to the first and so on, until the end of the set (or until the tie is reached). We aim to derive the probability of the event $E_\pla^{\alpha, \beta, \plc}$. We know that  this event is equivalent to a succession of independent (albeit not identically distributed)  binomial  experiments with parameters $(m, p_a)$ or $(m, 1-p_b)$. Hence all related distributions will beÊ relatively straightforward to derive as soon as  we are able to compute, for any given score, the number $K$, say, of complete service sequences (of $m$ rallies) and the number $R\,(<m)$, say, of remaining serves (by either $\pla$ or $\plb$) during the last service sequence.  This is performed in the following lemma (which we state without proof).  

\begin{lemma}\label{lemmasum}
Let  $0\le \alpha , \beta \le n-1$. Define the integers $K$ and $R$ as the unique solutions of   $\alpha+\beta=Km+R$ with $R\le m-1$. Then the event $E_\pla^{\alpha, \beta, \plc}$ can be decomposed into $k_1:=\lceil K/2\rceil$ service sequences by $\pla$ and $k_2:=\lfloor K/2\rfloor$ service sequences by $\plb$, ended by $R$ serves of $\pla$ (resp., of $\plb$) when $K$ is even (resp., when $K$ is odd). 
\end{lemma}

Let us now consider the event $E_A^{\alpha,\beta,C}$ for $\plc\in\{\pla, \plb\}$. From the nature of the game it is clear that, given the score $(\alpha,\beta)$, the only random quantity left is the number of points $\pla$ (or, equivalently,  $\plb$) has obtained on his/her own serve. Hence it is natural to define the events   $E_\pla^{\al,\be,\plc}(j) \subset E_\pla^{\al, \be, \plc}$ for which the scoring sequence occurs with $\pla$ scoring exactly $j$ points on his/her serve. 
Denoting by $p_{\pla}^{\al,\be,\plc}(j)$ the corresponding probability, these quantities can then be derived by exploring the essentially binomial nature of the experiment, as summarized in the following lemma.
\begin{lemma}\label{terrible}
Let $\al,\be\in\Nn$, and define $K,R$ as in Lemma \ref{lemmasum}. Also, for $\plc \in \{\pla, \plb\}$, define $\delta_\pla^\plc$ as the indicator of the equality between $\pla$ and $\plc$. Then, (i)  if  $R>0$ and $K$ is even (or $R=0$ and $K$ is odd),
$$p_{\pla}^{\al,\be,\plc}(j)=\displaystyle{\binom{\lceil{\frac{K}{2}}\rceil m+R-1}{j-\delta_\pla^\plc}p_a^j(1-p_a)^{\lceil{\frac{K}{2}}\rceil m+R-j}\binom{\lfloor{\frac{K}{2}}\rfloor m}{\alpha-j}(1-p_b)^{\alpha-j}p_b^{\lfloor{\frac{K}{2}}\rfloor m -\alpha+j}},$$
where $j\in \{\max(\delta_\pla^\plc,\alpha-\lceil{\frac{K}{2}}\rceil m),\ldots,\min(\alpha,\lceil{\frac{K}{2}}\rceil m+R-1+\delta_\pla^\plc)\}$;
(ii) if  $R>0$ and $K$ is odd  (or $R=0$ and $K$ is even),
$$p_{\pla}^{\al,\be,\plc}(j)=\displaystyle{\binom{\lceil{\frac{K}{2}}\rceil m}{j}p_a^j(1-p_a)^{\lceil\frac{K}{2}\rceil m-j}\binom{\lfloor{\frac{K}{2}}\rfloor m + R -1}{\alpha-j-\delta_\pla^\plc}(1-p_b)^{\alpha-j}p_b^{\lfloor{\frac{K}{2}}\rfloor m+R-\alpha+j}},$$
where $j\in \{\max(0,\alpha-\lfloor{\frac{K}{2}}\rfloor m-R+1-\delta_\pla^\plc),\ldots,\min(\alpha-\delta_\pla^\plc, \lceil{\frac{K}{2}}\rceil m)\}$.

\end{lemma}

Note that the above condensed formulae are easier to read if spelt out explicitly in each of the 8 possible cases. We therefore provide a complete version of Lemma~\ref{terrible} in the Appendix, accompanied by a formal proof. Summing up it is now a simple matter to obtain the final formulae -- at least for game scores that do not involve a tie.
\begin{theorem}\label{cool}
Fix $\al,\be\in\Nn$, and let $p_\pla^{\alpha, \beta, \plc}:={\rm P}[E_\pla^{\alpha, \beta, \plc}]$, where $E_\pla^{\alpha, \beta, \plc}:=\bigcup_{j}E_\pla^{\alpha, \beta, \plc}(j)$ with $\plc\in\{\pla, \plb\}$. Then $p_\pla^{\alpha,\beta,\pla}=\displaystyle{\sum_{j=0}^{\infty} p_{\pla}^{\al,\be,\pla}(j)}$ and $p_\pla^{\alpha,\beta,\plb}=\displaystyle{\sum_{j=0}^{\infty}p_{\pla}^{\al,\be,\plb}(j)}$, where the probabilities $p_\pla^{\alpha, \beta, \plc}(j)$ are defined in  Lemma~\ref{terrible} . 
\end{theorem}
When the parameters are chosen so as to satisfy the constraint $p_a=1-p_b=p$, the corresponding results concur with those obtained    for the no-server model. In all other cases, these results do not allow for an agreeable form so we dispense with their explicit expression. The behavior of these quantities in terms of the ruling parameters of the game is  illustrated (both in the old and new scoring systems) in Figure \ref{dist}.  \black
Finally regarding the winning probabilities in case of a tie, the same reasoning as for \eqref{super2} reveals that the probability that $\pla$ (resp., $\plb$) wins the set after $2\ell$ points is given by $p_a(1-p_b)((1-p_a)(1-p_b)+p_ap_b)^{\ell-1}$ (resp., by $(1-p_a)p_b((1-p_a)(1-p_b)+p_ap_b)^{\ell-1}$).   The probability of $\pla$ (resp., of $\plb$)  winning the set after a tie at $n-1$ readily follows and thus, combining Theorem \ref{cool}  and equations (\ref{pAA}) and (\ref{pAB}) yields $p_\pla^\pla$ and $p_\pla^\plb$, as desired. Because  these probabilities are particularly cumbersome to spell out explicitly we rather  choose to illustrate their behavior  (both in the old and new scoring systems) numerically and graphically (see Table \ref{tab:prob} and the different figures below). \black

\subsection{Distribution of the number of rallies in a single set.}\label{numbrall}

The results  established in the previous section allow us now to investigate the distribution of the number of rallies, $D$ say, needed to end a single set.  
Let $S$ stand for the random variable recording the first server of a set, and, for $\plc \in \{\pla, \plb\}$ and $E$ an event,  denote by $P_\plc[E] := {\rm P}[EÊ\, | \, S=\plc]$ the probability of $E$ conditional upon who serves first in a set. The main object of interest in this section is the mapping $d\mapsto P_\pla[D=d]$ for $d\in \Nn$.  
  
\begin{figure}[!]
\begin{center}\includegraphics[angle=0,width=1\linewidth]{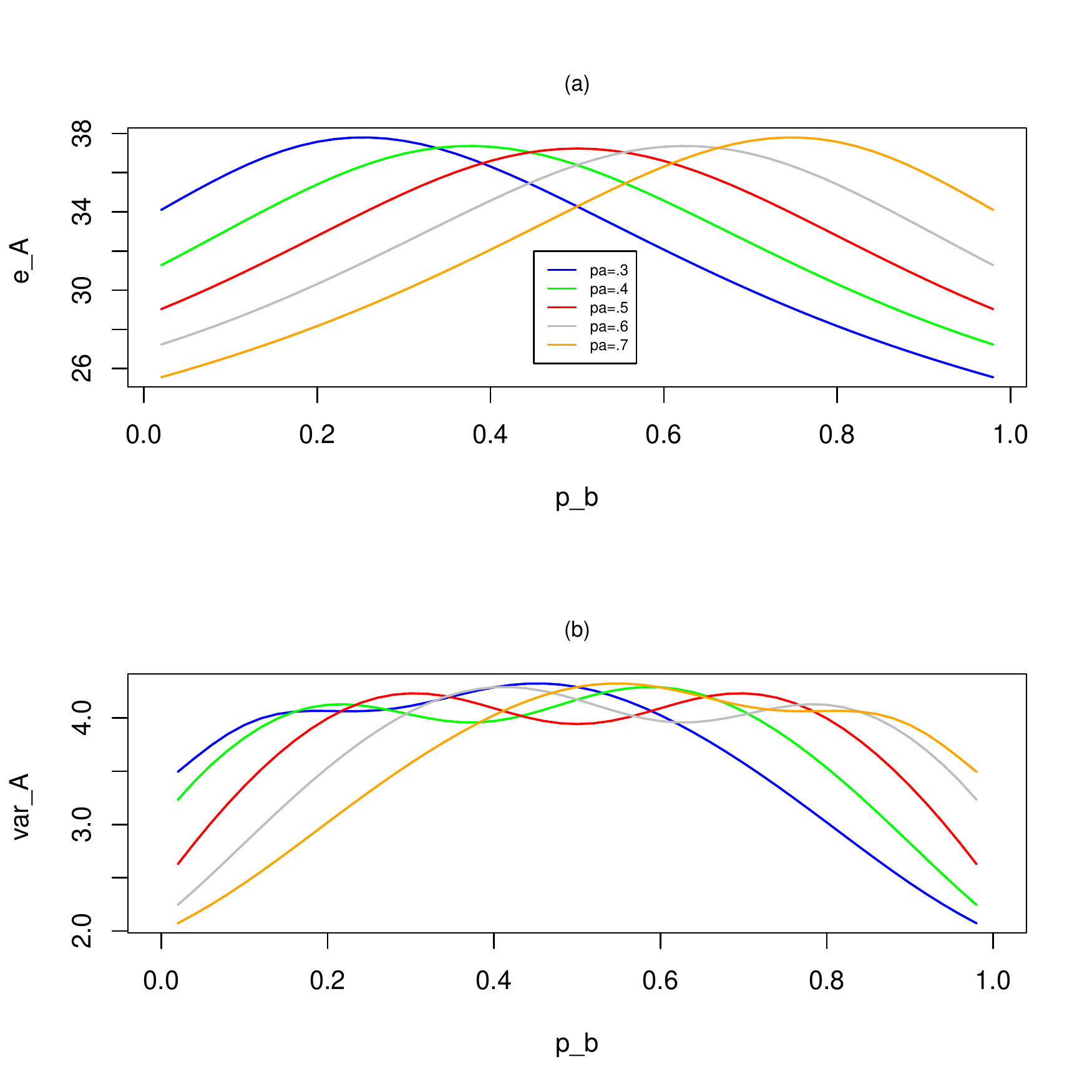}\end{center}
\caption{Evolution of expectation and variance} 
\label{evol} \footnotesize{Both subfigures refer to an $\pla$-set played under the $(5, 21, 1)$-scoring system. Subfigure (a) shows the evolution of the expectation and subfigure (b) shows the evolution of the variance of the duration $D$ unconditional on the winner of the set. }
\end{figure}

Obviously, because a point is scored on every rally, the number of rallies needed to reach a score $(\alpha, \beta)$ is  equal to $\alpha+\beta$. Hence  $P_\pla[D=d]=0$ for all $d\le n-1$ and $P_\pla[D=d]=p_\pla^{n, d-n, \pla}+p_\pla^{ d-n,n, \plb}$ for all $n\le d \le 2(n-1)$. Finally, for $d\ge 2n-1$, such length can only be achieved through the occurrence of a tie so that $P_\pla[D=d] = 0$ if $d=2n-1$ and   otherwise  $P_\pla[D=d] = (p_\pla^{n-1, n-1,\pla}+p_\pla^{n-1, n-1,\plb})  p_{\rm tie}^{d, n}$, where the latter quantity gives the probability that there are exactly $d-2(n-1)$ rallies after the tie. 
%
%
%
%
All that  remains is therefore  to compute $p_{\rm tie}^{d, n}$. For this first note  that $p_{\rm tie}^{d, n} = 0$ if $d-2(n-1)$ is odd. 
Next it is a simple matter of applying the same  arguments as in Section \ref{fund_res} to obtain  
$p_{\rm tie}^{d, n} = (p_a(1-p_b)+(1-p_a)p_b)((1-p_a)(1-p_b)+p_ap_b)^{\ell-1}$
with $2\ell = d-2(n-1)$. The distribution  of $D$ (as well as its moments) readily follows, as illustrated in Table~\ref{tab:prob},  Figure~\ref{evol}  and  Figure~\ref{distD}.

The evolution of the expectation and variance of the number of rallies $D$ in an $\pla$-set unconditional on the winner of that set are displayed in Figure~\ref{evol}.  (Note that the effect of a variation of $n$ in one set played according to the $(n, m, G)$-rule will be a straightforward rescaling. Plots are therefore not provided.)
The expectation curves are quite smooth and unimodal around $p_a=p_b$, while the variance curves are also smooth but exhibit a slightly bimodal nature around $p_a=p_b$. These findings are in accordance with the intuition that players of approximately the same strength play longer. Moreover, when the strengths are only slightly different, the variance becomes biggest, as in that case there may be both tight and uneven sets. Finally note that when $p_b\rightarrow0$ or $p_b\rightarrow1$ the expected duration is not equal to 21 as one might think at first sight but is rather around 25. This evidently follows from the fact that $p_a$ is confined between .3 and .7, and hence when player $\pla$ serves the outcomes remain random.  


\section{Comparing   scoring systems in table tennis.}\label{compare}

 In this section, we discuss the influence of the parameters $(m, n, G)$  in terms of score distributions, durations and match-winning probabilities. Our main interest is to discuss the effects of the switch from the ``old'' $(5,21,G)$-scoring system to the ``new'' $(2,11,G)$-scoring system. 

\begin{figure}[!]
\begin{center}\includegraphics[angle=0,width=1\linewidth]{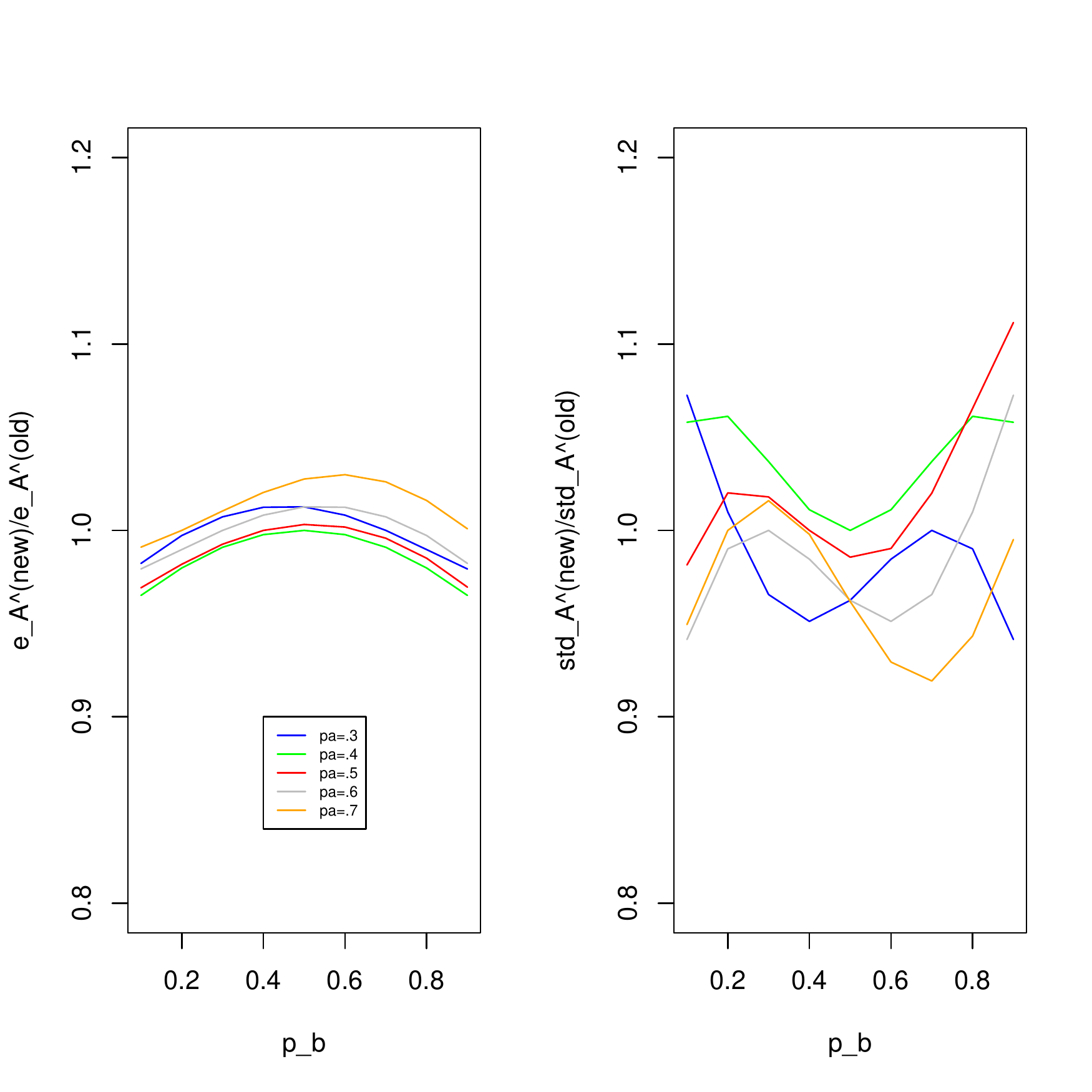}\end{center}
\caption{Ratio of expectations and standard deviations, as a function of $p_b$, between new and old scoring systems} 
\label{ratio_m} \footnotesize{Both plots refer to an $\pla$-set played with $n=11$ and $G=1$ for different values of $p_a$, as a function of $p_b$. Plot (a) shows the ratio between the expected durations under the $(11, 2, 1)$ rule and that under the $(11, 5, 1)$ rule; Plot (b) reports the ratio between the standard deviations. }
\end{figure}

We first stress how little influence the choice of $m$ has on the different distributions, as long as $m$ is chosen so as to satisfy the above described divisibility constraints.  This is illustrated in Figure \ref{ratio_m}, where the ratio between the expected durations  (Figure \ref{ratio_m} a) and standard deviations (Figure \ref{ratio_m} b) under the $(5, 11, 1)$ and the $(2, 11, 1)$ rules are reported. Aside from the fact that the different curves cross the horizontal line $y=1$ at each couple $(p_a, p_b)$ such that $p_a+p_b=1$, we draw the reader's attention to how little variation these quantities endure. This encourages to disregard the role of $m$ in the future, and concentrate on the dependence on the parameters $n$, $G$ and $(p_a, p_b)$.

\begin{table}[!]
\begin{scriptsize}
\begin{center}
\caption{Probability and Expected Duration} \label{tab:prob}
\begin{tabular}{c|ccccccccc}
\hline \hline
   & \multicolumn{9}{c}{$p_b$}\\
   & 0.1 & 0.2 & 0.3 & 0.4 & 0.5 & 0.6 & 0.7 & 0.8 & 0.9 \\
\hline
 $p_a$   & \multicolumn{9}{c}{Probability that $\pla$ wins a set when $\pla$ serves first with $m=5$ and $n=21$}\\
0.1 & 0.5000 &  0.1657& 0.0404 & 0.0072 & 0.0008  & 0 & 0 & 0  & 0 \\
0.2 &0.8342 & 0.5000 & 0.2190 & 0.0702 & 0.0159 & 0.0023 & 0.0002 & 0 & 0 \\
0.3 & 0.9595& 0.7809 & 0.5000 & 0.2430 & 0.0863 & 0.0209 & 0.0030 & 0.0001 & 0 \\
0.4 & 0.9928 & 0.9297 & 0.7569 & 0.5000 & 0.2530 & 0.0914 & 0.0209 & 0.0023 & 0 \\
0.5 & 0.9991 & 0.9840 & 0.9136 & 0.7469 & 0.5000 & 0.2530 & 0.0863 & 0.0159 & 0.0008 \\
0.6 & 0.9999 & 0.9976 & 0.9790 & 0.9085 & 0.7469 & 0.5000 & 0.2430 & 0.0702 & 0.0072 \\
0.7 & 0.9999 & 0.9998 & 0.9969 & 0.9790 & 0.9136 & 0.7569 & 0.5000 & 0.2190 & 0.0404 \\
0.8 & 1 & 0.9999 & 0.9998 & 0.9976 & 0.9840 & 0.9297 & 0.7809 & 0.5000 & 0.0166 \\
0.9 & 1 & 1 & 0.9999 & 0.9999 & 0.9991 & 0.9928 & 0.9595 & 0.8342 & 0.5000 \\
\hline
  $p_a$  & \multicolumn{9}{c}{Probability that $\pla$ wins a set when $\pla$ serves first with $m=2$ and $n=11$}\\
0.1 & 0.5000 &  0.2300& 0.0951 & 0.0343 & 0.0103  & 0.0024 & 0.0003 & 0  & 0 \\
0.2 &0.7699 & 0.5000 & 0.2819 & 0.1376 & 0.0568 &  0.0189 & 0.0046 & 0.0006 & 0 \\
0.3 & 0.9048 & 0.7180 & 0.5000 & 0.3035 & 0.1576 & 0.0673 & 0.0219 & 0.0046 & 0.0003 \\
0.4 &0.9656 & 0.8623 & 0.6964 & 0.5000 & 0.3121 & 0.1635 & 0.0673 & 0.0189 & 0.0024 \\
0.5 & 0.9896 & 0.9431 & 0.8423 & 0.6878 & 0.5000 & 0.3121 & 0.1576 & 0.0568 & 0.0103 \\
0.6 & 0.9975 & 0.9810 & 0.9326 & 0.8364 & 0.6878 & 0.5000 & 0.3035 & 0.1376 & 0.0343 \\
0.7 & 0.9996 & 0.9953 & 0.9780 & 0.9326 & 0.8423 & 0.6964 & 0.5000 & 0.2819 & 0.9513 \\
0.8 & 0.9999 & 0.9993 & 0.9953 & 0.9810 & 0.9431 & 0.8623 & 0.7180 & 0.5000 & 0.2300 \\
0.9 & 0.9999 & 0.9999 & 0.9996 & 0.9975 & 0.9896 & 0.9656 & 0.9048 & 0.7699 & 0.5000 \\
\hline
 $p_a$   & \multicolumn{9}{c}{Expected duration of a set when $\pla$ serves first with $m=5$ and $n=21$}\\
0.1 & 40.1616 &  36.5982 & 33.7171 & 31.3842 & 29.3013  & 27.4362 & 25.8432 & 24.5067  & 23.3333 \\
0.2 & 38.6311 & 38.2747 & 36.3666 & 34.0094 & 31.7268 & 29.6526 & 27.8216 & 26.2500 & 24.9032 \\
0.3 & 36.0125& 37.5782 & 37.6279 & 36.3163 & 34.2761 & 32.0696 & 29.9949 & 28.1681 & 26.6186 \\
0.4 & 33.1567 & 35.4021 & 36.9720 & 37.3272 & 36.3937 & 34.5748 & 32.4159 & 30.3135 & 28.4663 \\
0.5 & 30.5940 & 32.7743 & 34.9382 & 36.6022 & 37.2359 & 36.6022 & 34.9382 & 32.7743 & 30.5940 \\
0.6 & 28.4663 & 30.3135 & 32.4159 & 34.5748 & 36.3937 & 37.3272 & 36.9720 & 35.4021 & 33.1567 \\
0.7 & 26.6186 & 28.1681 & 29.9949 & 32.0696 & 34.2761 & 36.3163 & 37.6279 & 37.5782 & 36.0125 \\
0.8 & 24.9032 & 26.2500 & 27.8216 & 29.6526 & 31.7268 & 34.0094 & 36.3666 & 38.2747 & 38.6311 \\
0.9 & 23.3333 & 24.5068 & 25.8433 & 27.4362 & 29.3013 & 31.3843 & 33.7172 & 36.5982 & 40.1616 \\
\hline
  $p_a$  & \multicolumn{9}{c}{Expected duration of a set when $\pla$ serves first with $m=2$ and $n=11$}\\
0.1 & 22.5402 &  19.9063& 18.0171 & 16.5767 & 15.4215  & 14.4575 & 13.6267 & 12.8901  & 12.2222 \\
0.2 & 20.4333 & 20.0069 & 19.0096 & 17.8241 & 16.6473 &  15.5657 & 14.6029 & 13.7513  & 12.9904 \\
0.3 & 18.7138 & 19.3163 & 19.2465 & 18.6612 & 17.7620 & 16.7301 & 15.6955 & 14.7315 & 13.8635 \\
0.4 & 17.2440 & 18.2349 & 18.8266 & 18.9228 & 18.5459 & 17.8052 & 16.8511 & 15.8300 & 14.8509 \\
0.5 & 15.9686 & 17.0222 & 17.9547 & 18.5985 & 18.8285 & 18.5985 & 17.9547 & 17.0222 & 15.9686 \\
0.6 & 14.8509 & 15.8300 & 16.8511 & 17.8052 & 18.5459 & 18.9228 & 18.8266 & 18.2349 & 17.2440 \\
0.7 & 13.8635 & 14.7315 & 15.6955 & 16.7301 & 17.7620 & 18.6612 & 19.2465 & 19.3163 & 18.7138 \\
0.8 & 12.9904 & 13.7513 & 14.6029 & 15.5657 & 16.6473 & 17.8241 & 19.0096 & 20.0069 & 20.4333 \\
0.9 & 12.2222 & 12.8901 & 13.6267 & 14.4575 & 15.4215 & 16.5767 & 18.0171 & 19.9063 & 22.5402 \\
\hline \hline
\end{tabular}
\end{center}
\end{scriptsize}
\end{table}

Table~\ref{tab:prob} shows the probability that $\pla$ wins a set, in which he/she serves first, both for the old and new scoring system, as well as the expected duration of such a set in both scoring systems. The two first parts in Table~\ref{tab:prob} list the values of $p_\pla^\pla$ for all possible combinations of couples $(p_a,p_b)$ varying respectively from 0.1 to 0.9 with a step size of 0.1. A simple inspection of the probabilities indicates that the stronger one of the two players inevitably wins the set, which is a confirmation of the common intuition. The main difference between the two scoring systems lies in the  fact that on the one hand, when $p_a>p_b$ (values under the diagonal), player~$\pla$ has a higher chance to win the set in the old system than in the new one, whereas on the other hand, when $p_a<p_b$ (values over the diagonal), player~$\pla$'s (small) chances to win the set are higher in the new system. This can be explained as follows: the more points are necessary to win the set, the more important the strengths $p_a$ and $p_b$ become, and the ``victory against the odds'' is therefore less probable in the old scoring system than in the new one. We also draw the reader's attention to the fact that $p_\pla^\pla(p_a,p_b)=p_\pla^\pla(1-p_b,1-p_a)$ and $p_\pla^\pla(p_a,p_b)+p_\pla^\pla(p_b,p_a)=1$ (with $p_\pla^\pla(x,y)$ standing for the probability that player $\pla$ wins an $\pla$-set under the conditions $p_a=x$ and $p_b=y$, $x,y\in(0,1)$). These phenomena are discussed  in Schulman and Hamdan (1977), and will not be further elaborated on here.

\begin{figure}[!]
\begin{center}\includegraphics[angle=0,width=1\linewidth]{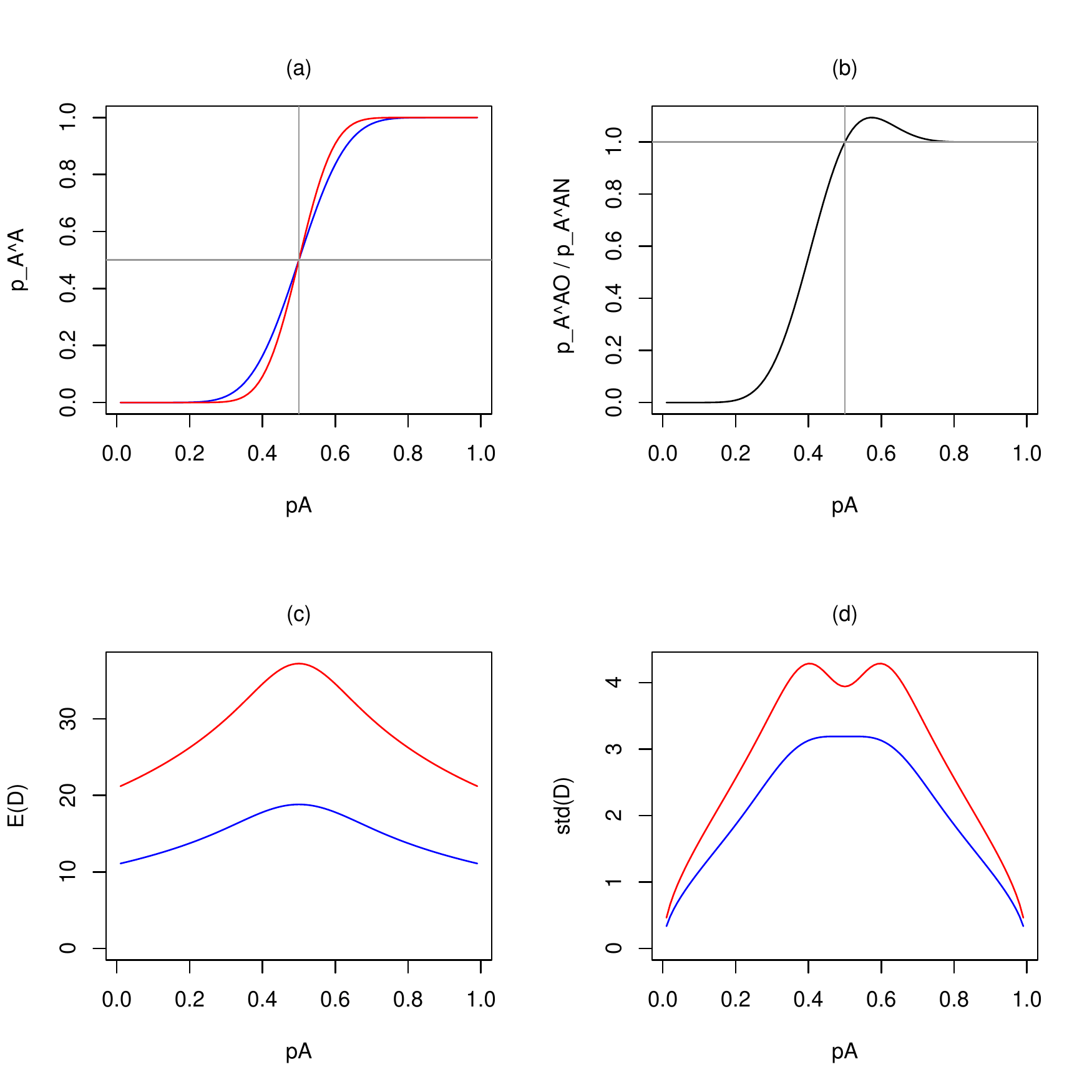}\end{center}
\caption{Comparison: one set} 
\label{compset} \footnotesize{As a function of $p=p_{a}=1-p_{b}$ (that is, in the no-server model), subfigure (a) contains the probabilities $p_{\pla}^{\pla}$ (in red) that player $\pla$ wins an $\pla$-set in the (21, 5, 1)-scoring system, along with the probabilities $p_{\pla}^{\pla}$ (in blue) that player $\pla$ wins an $\pla$-set in the (11, 2, 1)-scoring system.   Subfigures (b) and (c) show the expectation and the standard deviation of the number of rallies needed to complete the corresponding set, unconditional on the winner.}
\end{figure}

\begin{figure}[!]
\begin{center}\includegraphics[angle=0,width=1\linewidth]{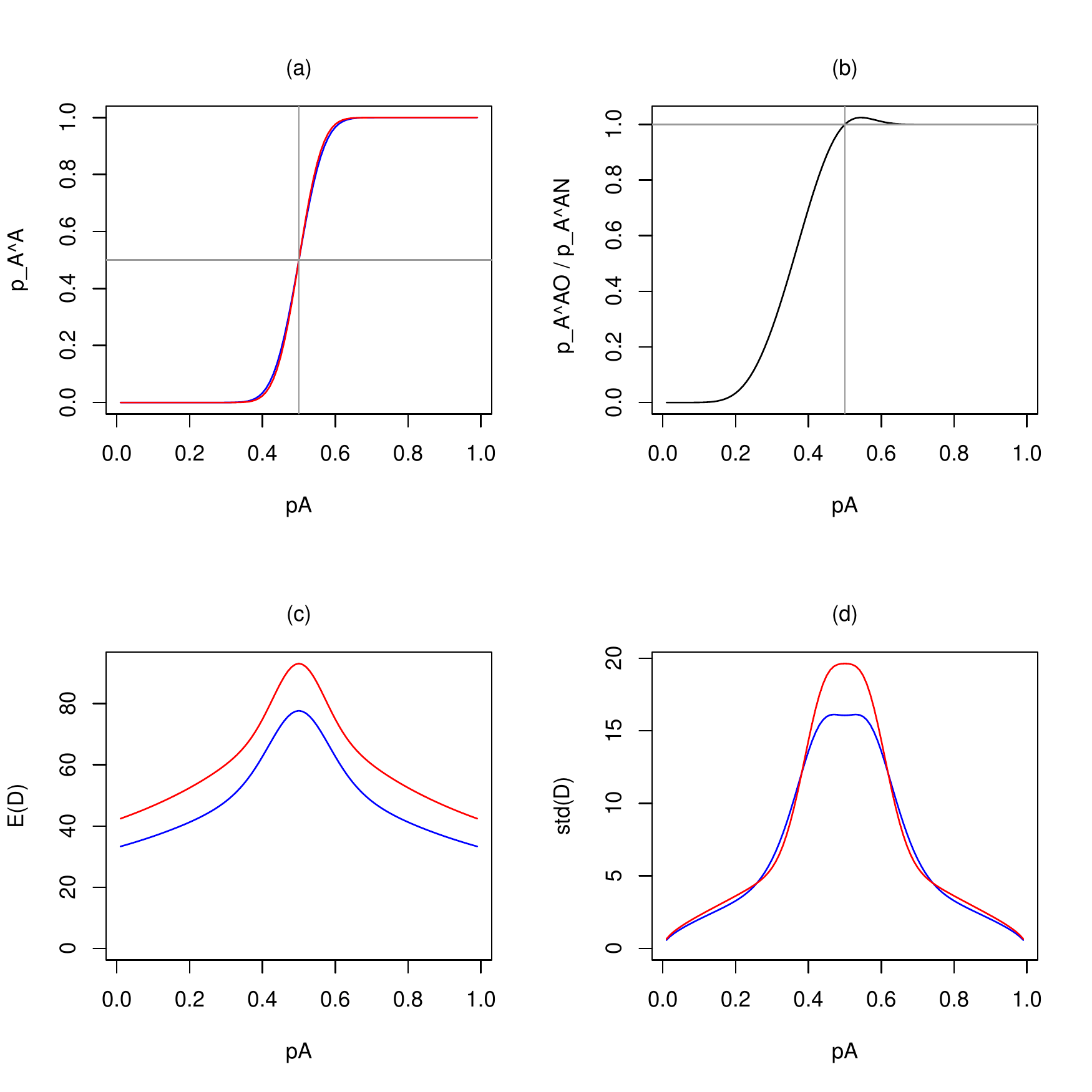}\end{center}
\caption{Comparison: a match} 
\label{compmatch} \footnotesize{As a function of $p=p_{a}=1-p_{b}$ (that is, in the no-server model), subfigure (a) contains the probabilities $p_{\pla}^{\pla}$ (in red) that player $\pla$ wins an $\pla$-match in the (21, 5, 3)-scoring system, along with the probabilities $p_{\pla}^{\pla}$ (in blue) that player $\pla$ wins an $\pla$-match in the (11, 2, 4)-scoring system.   Subfigures (b) and (c) show the expectation and the standard deviation of the number of rallies needed to complete the corresponding match, unconditional on the winner.}
\end{figure}

\begin{figure}[!]
\begin{center}\includegraphics[angle=0,width=1\linewidth]{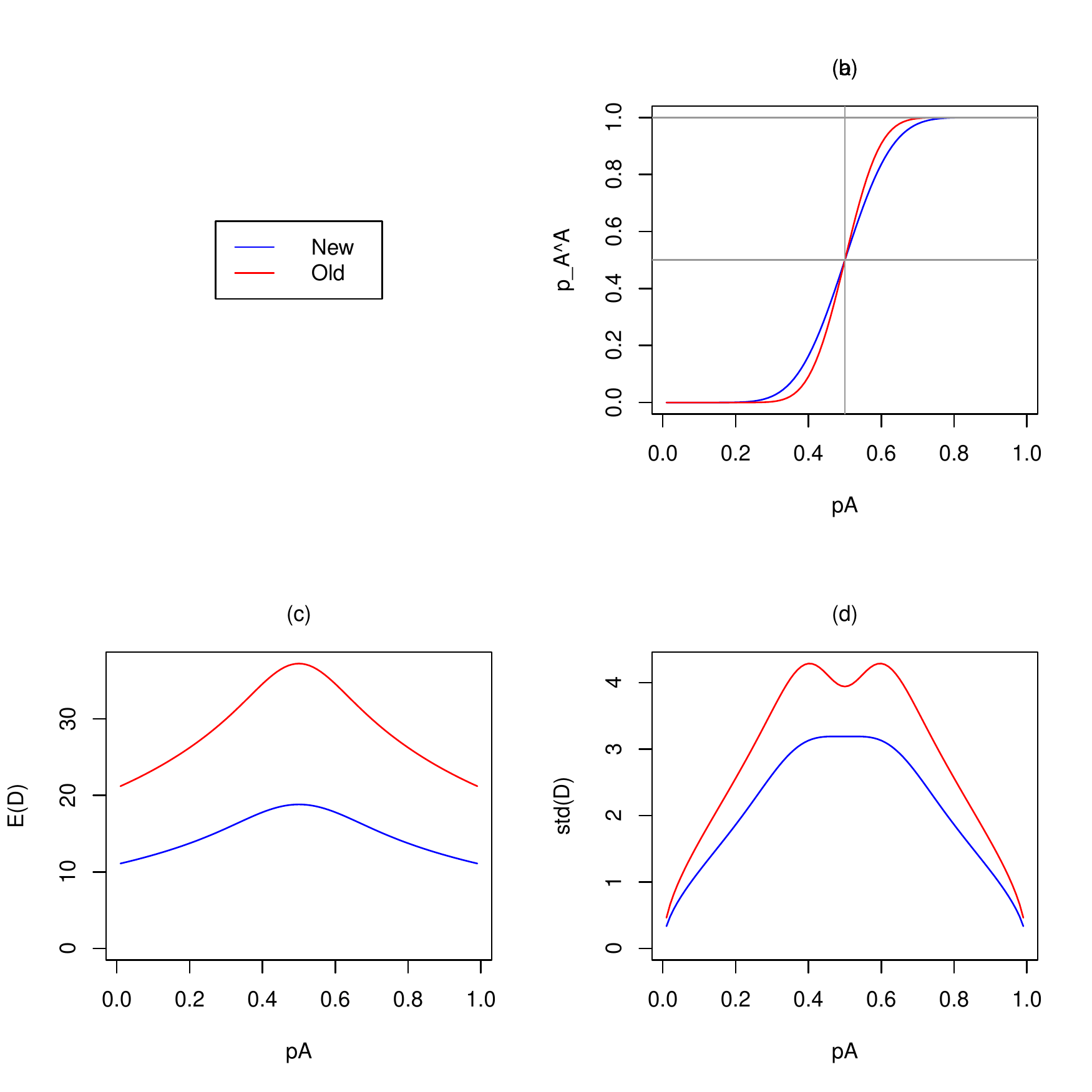}\end{center}
\caption{Comparison: a match} 
\label{compmatch2} \footnotesize{As a function of $p=p_{a}=1-p_{b}$ (that is, in the no-server model), subfigure (a) contains the probabilities $p_{\pla}^{\pla}$ (in red) that player $\pla$ wins an $\pla$-match in the (21, 5, 4)-scoring system, along with the probabilities $p_{\pla}^{\pla}$ (in blue) that player $\pla$ wins an $\pla$-match in the (11, 2, 8)-scoring system.   Subfigures (b) and (c) show the expectation and the standard deviation of the number of rallies needed to complete the corresponding match, unconditional on the winner.}
\end{figure}

The two last parts of Table~\ref{tab:prob} show the expected duration of a set in which player $\pla$ serves first. The sets with the largest expected durations are those on the diagonal, where both players are roughly of similar strength. Clearly the duration of a set is smaller in the new scoring sytem than in the older one. More interesting is the fact  that the duration of a set in the old scoring system is always roughly  double that in the new scoring system or, more precisely, that we have  $e_\pla^{\pla, old}/e_\pla^{\pla, new}  \approx 21/11$. This seemingly linear relationship between the average length of a set and the number of points played is easily confirmed by further computations, as  shown in Table \ref{tab:prob2}.  Likewise, from Table \ref{tab:prob2}, we see that the same multiplicative factor (under a square root) appears when considering the standard deviation of $D$.   The same conclusions hold true when passing from a single set to a full match.  From this we deduce that  in order for the ITTF, for instance,  to derive a scoring triple $(m_2, n_2, G_2)$ which \emph{does not} change the average length of the game when switching from the original triple $(m_1, n_1, G_1)$, it is advisable to choose the  number of sets $G_2$ in such a way that   $G_1n_1 = G_2n_2$. If, for instance, $n_1 = 21$ and $G_1 = 3$, then the choice of $n_2=11$ and $G_2 = 6$ would influence the expected durations less than $G_2 = 3$ or $G_2=4$. The choice of $G_2=3$ or 4 nevertheless ensures a difference that is negligible (see Figures  \ref{compset} and  \ref{compmatch} and \ref{compmatch2} where different choices of triples are compared in the no-server model).

\begin{table}[!]
\begin{scriptsize}
\begin{center}
\caption{Ratio of  Expected Durations and Standard Deviations}\label{tab:prob2} 
\begin{tabular}{c|ccccc}
\hline \hline
$n_2/n_1$    & 0.1 & 0.2 & 0.3 & 0.4 & 0.5 \\

\hline
   & \multicolumn{5}{c}{Ratio of expectations}\\
  
$ 21/11 = 1.9090$ & 1.9000 &  1.9058 & 1.9297 & 1.9623 & 1.9776   \\
$ 31/21 = 1.4761$ & 1.4670 &  1.4697 & 1.4797 &  1.4963 & 1.5052   \\
$ 41/31 = 1.3225$ & 1.3149 &  1.3172 & 1.3221 &  1.3326 & 1.3392 \\
$ 31/11 = 2.8181$ & 2.8231  & 2.8276 & 2.8728 & 2.9419 &  2.9769 \\
$ 41/11 = 3.7272$ & 3.7482  & 3.7527 & 3.8151 & 3.9260 &  3.9867 \\
\hline
   & \multicolumn{5}{c}{Ratio of Standard Deviations }\\
$\sqrt{21/11} = 1.3816 $  & 1.3437 &  1.3037 & 1.3426 & 1.2975  &  1.2356 \\
$ \sqrt{ 31/21} = 1.2149 $  & 1.1789 &  1.2171 & 1.2527 &  1.2356 & 1.1754  \\
$ \sqrt{ 41/31} = 1.1500 $ & 1.1553 &  1.1500 & 1.1790 &  1.1863 & 1.1349 \\
$ \sqrt{ 31/11} = 1.6787 $  & 1.5490  & 1.6080 & 1.6819 & 1.5944 &  1.4524 \\
$ \sqrt{ 41/11} = 1.9306 $  & 1.8476 & 1.8529 & 1.9749 & 1.8816 &  1.6484 \\
\hline \hline
\end{tabular}
\end{center}
\end{scriptsize}
\end{table}


\begin{figure}[!]
\begin{center}\includegraphics[angle=0,width=1\linewidth]{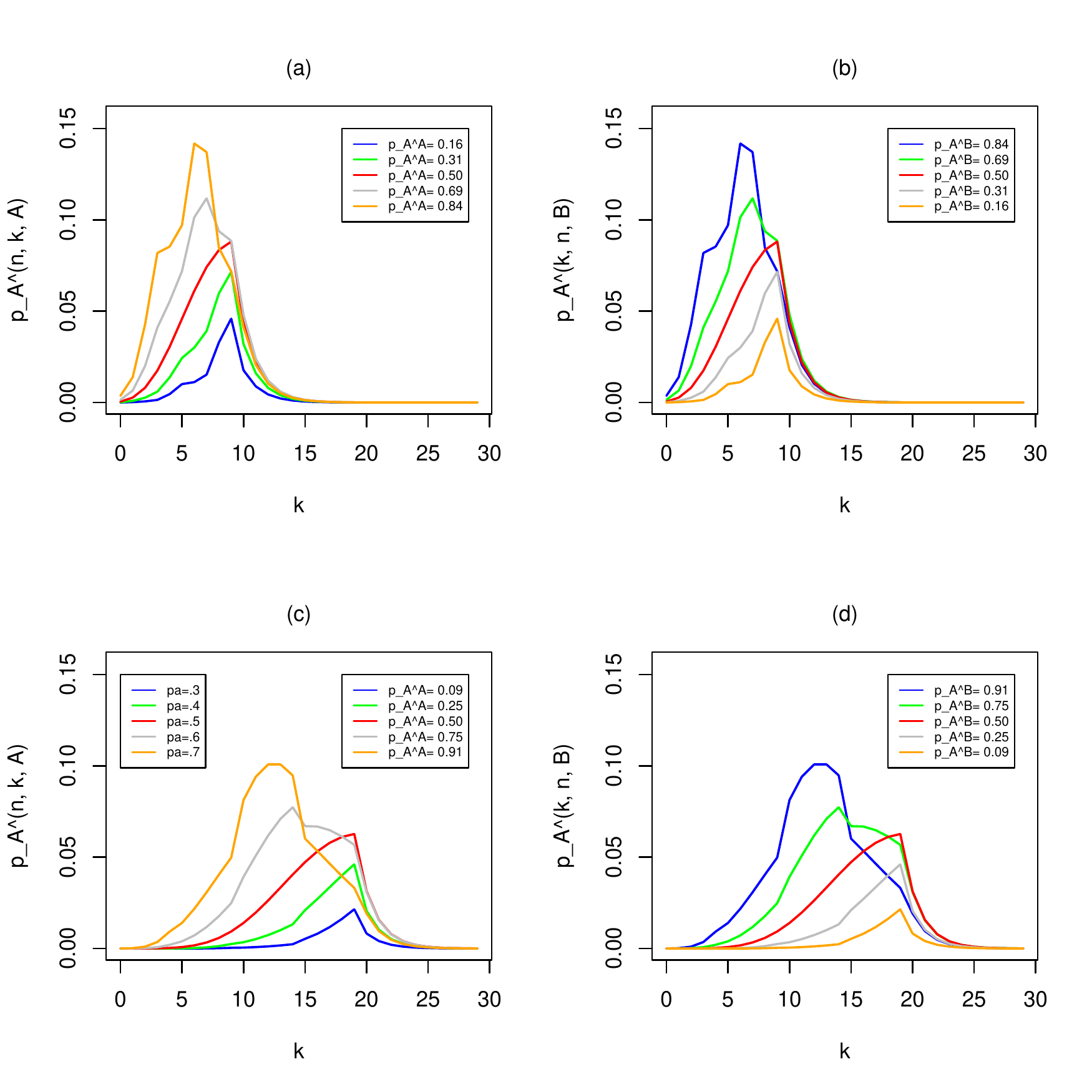}\end{center}
\caption{Distribution of the scores with $p_b=.5$} 
\label{dist} \footnotesize{Subfigures (a) and (b) refer to an $\pla$-set played under the $(2, 11, 1)$-scoring system. Subfigure (a): for $(p_{a},p_{b})=(.7,.5),(.6,.5),(.5,.5),(.4,.5)$ and $(.3,.5)$, probabilities $p_{A}^{n,k,A}$ that player $\pla$ wins the set on the score $(n,k)$. Subfigure (b): the corresponding values for victories of $\plb$. Subfigures (c) and (d) refer to an $\pla$-set played under the $(5, 21, 1)$-scoring system. Subfigure (c): for $(p_{a},p_{b})=(.7,.5),(.6,.5),(.5,.5),(.4,.5)$ and $(.3,.5)$, probabilities $p_{A}^{n,k,A}$ that player $\pla$ wins the set on the score $(n,k)$. Subfigure (d): the corresponding values for victories of $\plb$. The different set-winning probabilities $p_\pla^\plc$ for $C\in\{\pla,\plb\}$ and each couple of values $(p_{a},p_{b})$ are given in each subfigure.
}
\end{figure}

Figure~\ref{dist} presents, for $n=11$ and $n=21$ respectively, the score distribution associated with $(p_{a},p_{b})=(.7,.5),(.6,.5),(.5,.5),(.4,.5)$ and $(.3,.5)$. Subplots (a) and (c) contain the probabilities that player $\pla$ wins an $\pla$-set in which the opponent scores a total of $k$ points (for instance, in the new scoring system, $k=0,\ldots,9$ corresponds to a victory of $\pla$ on the score $(11,0)$,\ldots,$(11,9)$, and in case of a tie, for which $k\geq10$, player~$\pla$ wins on scores of the form $(k+2,k)$) and the subplots (b) and (d) give the probability for $\plb$ winning an $A$-set where this time player~$\pla$ scores $k$ points. It appears that these distributions are highly sensitive to the values of $(p_{a},p_{b})$; obviously, this extends to the corresponding match-winning probabilities. In the $(5,21,G)$-scoring system, $p_{\pla}^{\pla}$ varies between $.09$ and $.91$ and $p_{\pla}^{\plb}$ between $.91$ and $.09$, when, for fixed $p_{b}=.5$, $p_{a}$ ranges from $.3$ to $.7$. For the other scoring system, $p_{\pla}^{\pla}$ takes values between $.16$ and $.84$, while $p_{\pla}^{\plb}$ ranges from $.84$ to $.16$. This confirms our previous findings that the difference in strengths is more important in the old than in the new system. The following three comments hold for both scoring systems: (i)~as~$\pla$ (resp., $\plb$) is getting stronger with respect to $\plb$ (resp., $\pla$), his/her winning probability dramatically increases, (ii) whatever the value of $p_a$, white-washes seldom occur, and (iii) the score distributions are not monotone in $p_a$. Finally note that subfigures~(a) and (b), as well as subfigures~(c) and (d), display the same curves with the colors in  reverse order; this is due to the symmetric roles played by the two players.

\begin{figure}[!]
\begin{center}\includegraphics[angle=0,width=1\linewidth]{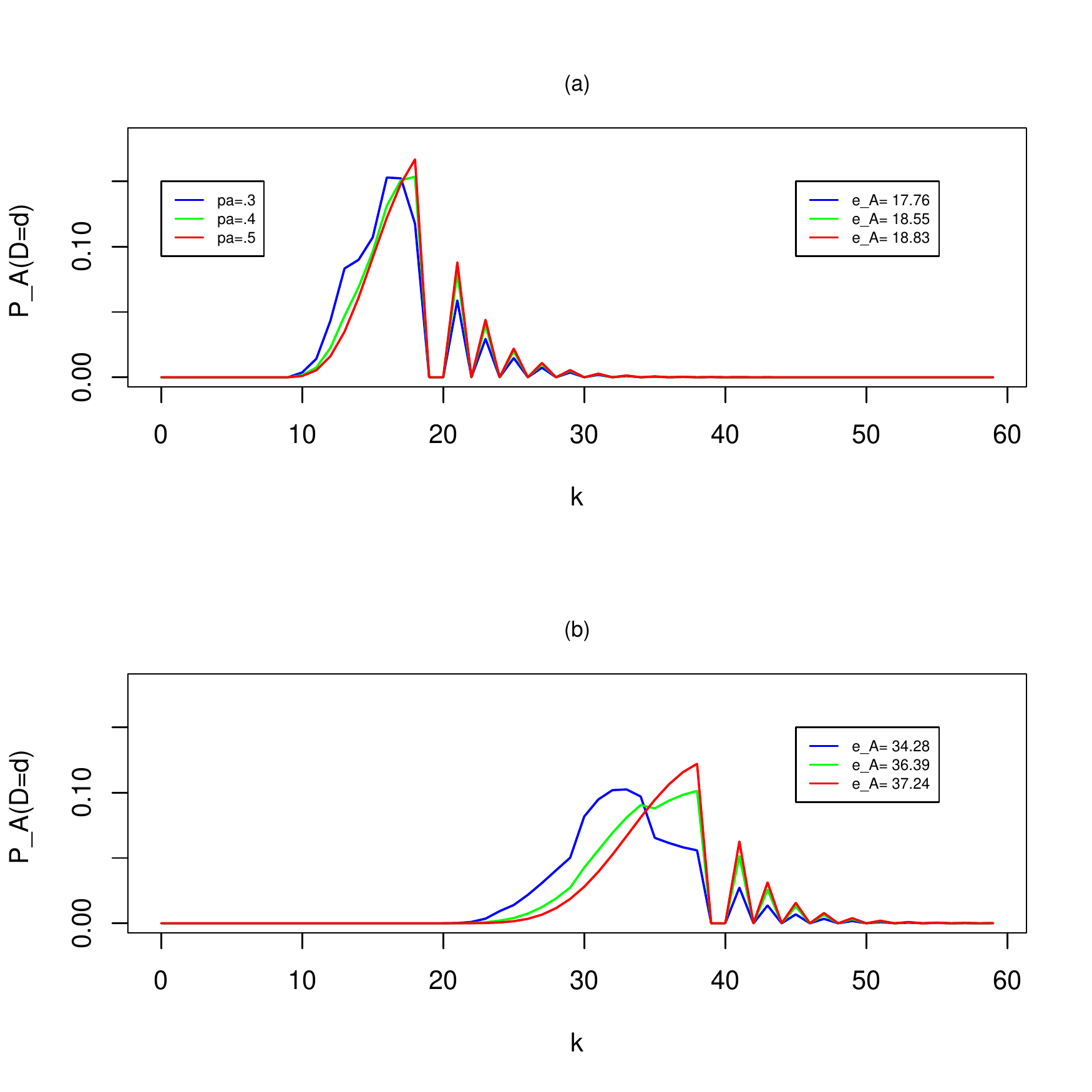}\end{center}
\caption{Distribution of $D$ with $p_b=.5$} 
\label{distD} \footnotesize{Both plots refer to an $\pla$-set. For $(p_{a},p_{b})=(.5,.5),(.4,.5)$ and $(.3,.5)$ they report the probabilities that the number of rallies $D$ needed to finish the set takes value $d$ unconditional on the winner of the set. Plot (a) shows the probabilities and expectations for the $(2,11,1)$-scoring system and plot (b) for the $(5,21,1)$-scoring system.}
\end{figure}

In Figure~\ref{distD} we plot, in the setup of an $\pla$-set and for $(p_{a},p_{b})=(.5,.5),(.4,.5)$ and $(.3,.5)$ (we omit the couples $(.7,.5)$ and $(.6,.5)$ as the symmetry of the situation implies a confounding with the plots for $(.3,.5)$ and $(.4,.5)$, respectively), the unconditional (on the winner of the set) probabilities that the number of rallies $D$ equals $d\in[0,60]$ with respect to that value $d$. The expectations of $D$ in each case are indicated as well. Subfigure~(a) corresponds to the new scoring system, and subfigure~(b) to the old one. In the old scoring system the probabilities are more dispersed than in the new scoring system; this is a direct consequence of the  larger number of  points needed to play until either player wins the set. For both plots, the peaks that occur after the main peak are due to the fact that a set can only end after an even number of rallies in case of a tie. The red curve is the one that takes the highest values, which confirms the intuition that two players of the same strength take more time to battle out the set and that the scores are quite tight (see also the related expected values $e_\pla=18.83$ in the new scoring system and $e_\pla=37.24$ in the old one).
 
The dependence of the set-winning probabilities on ($p_{a},p_{b}$) is of primary importance. In what follows, we will investigate this dependence visually by comparing both scoring systems in the no-server model ($p=p_{a}=1-p_{b}$). The results are illustrated in Figure~\ref{compset}.  Figure~\ref{compset}~(a) shows the probability that player $\pla$ wins an $\pla$-set as a function of his/her strength~$p$. The blue curve gives the probability for the new scoring system and the red curve for the old scoring system. The plot supports the claim that, for most values of $p$, the choice of the scoring system does not influence the set-winning probabilities. The only slight differences appear within $[.3,.45]$ and $[.55,.7]$, where the new scoring system quicker equalizes both players, which is in agreement with our previous findings. These observations are further supported by Figure~\ref{compset}~(b), where the ratio between the old and the new scoring system is plotted as a function of $p$. This graph reveals that although the probability that player $\pla$ wins an $\pla$-set is essentially the same for both scoring systems if he/she is the best player ($p_{\pla}^{\pla, \rm old}/p_{\pla}^{\pla, \rm new}\approx 1$ for $p>.7$, with evident notations),  for $p\in[.5,.7]$ the probability that $\pla$ wins the $\pla$-set in the old scoring system is slightly higher than in the new one, and  when $p<.5$, player~$\pla$ is more likely to win the set in the new system than in the old one. Finally, the ratio approaches~$0$ as $p$ tends to $0$ (for $p<.2$, say). This last remark indicates that the old scoring system leaves severely less winning chances to~$\pla$ when he/she is weak than the new one; in other words, we see again that the new scoring system appears more balanced. The duration plots simply confirm what common-sense already tells us. Indeed, we see in  Figure~\ref{compset}~(c) that the expectation of $D$ is, uniformly in $p\in(0,1)$, smaller for the new scoring system than for the old one; this is of course a direct consequence of the dramatic reduction of the number of points necessary to win the set. The same conclusion happens to be true for the standard deviation of $D$ (the old scoring system varies between $3.26$ and $4.29$, the new scoring system between $2.42$ and $3.22$), as illustrated in Figure~\ref{compset}~(d). The twin-peak shape of both standard error curves is most certainly due to the quasi-equality in rally-winning probabilities between both players which leads to both tighter final scores and more uneven scores. Finally note that both curves of Figures~\ref{compset}~(c) and (d) are symmetric about $p=.5$, as expected.

 We conduct, in Figure~\ref{compmatch}, a similar comparison between both scoring systems at match level. More precisely, we compare a $(5,21,3)$-scoring system (red curve) with a $(2,11,4)$-scoring system (blue curve). Subfigures (a) and (b) of Figure~\ref{compmatch} (to be compared with the corresponding subfigures in  Figure~\ref{compset}) indicate  that, for any fixed $p$,  the higher number of sets needed to achieve victory in the new scoring system ensures that  match-winning probabilities remain roughly the same before and after the rule change. Subfigures (c) and (d), i.e. the expectation and standard deviation plots, convey the same message as for a single set, except that here the two curves of the standard deviation cross at some points. Note that the standard deviation is in general smaller in the new scoring system, hence the change has made the length of the match more predictable. Finally in  Figure~\ref{compmatch2} we show, for the sake of illustration,  how  choosing $G_{\mbox{\small new}}$ and $G_{\mbox{\small old}}$ such that $n_{\mbox{\small new}}G_{\mbox{\small  new}} \approx n _{\mbox{\small old}}G_{\mbox{\small old}}$ ensures -- as is intuitively clear -- that both the probabilities and the expectations remain largely unchanged when switching from  $(m_{\mbox{\small old}}, n_{\mbox{\small old}}, G_{\mbox{\small old}})$ to $(m_{\mbox{\small new}}, n_{\mbox{\small new}}, G_{\mbox{\small new}})$.  

\section{Final comments.}\label{fc}

This paper provides a \emph{complete} probabilistic description  of games played according to the $(m, n, G)$-scoring system. It complements and extends the previous contribution by Schulman and Hamdan (1977). The formulae provided, although cumbersome, can easily be implemented for numerical purposes, and yield some striking illustrations. The results of our paper can be useful, in practice, to the sport community (e.g. the ITTF) and to TV broadcasting programs, as they allow to analyze any $(m,n,G)$-scoring system. Our findings have allowed us to perform an in-depth comparison of the old and new scoring system in table tennis.  If the aims behind the system change were (i) a better control of the length of a match and (ii) an increase in the potential number of crucial points without influencing the relative strengths of the players too much (i.e. without changing the winning probabilities),  then our results confirm  that these goals are achieved. 

We also mention how our results  also allow for constructing elegant estimation procedures of the respective strengths of the players. Obviously disposing of the full formulas permits to work out Maximum Likelihood Estimators for $p_a$, $p_b$. These will necessarily be obtained by numerical maximization on basis of the formulae in Lemma~\ref{terrible}. There is, however, a more elegant   (and efficient) way to perform such estimation, which only requires retaining one more information from a given encounter between two players, namely the 
number of points scored by player $\pla$ (resp., player $\plb$) on his/her own serve and on the serve of his/her opponent. Indeed with this information one can show that  the maximum likelihood estimator for $p_a$, say, is given by the  ratio of the number of points won by $\pla$ on his/her serve and  the total number of his/her serves.  Based on Bradley-Terry paired comparison methods, these estimators will allow for constructing   interesting new ranking methods within round-robin tournaments. This will be the subject of a future publication.


Finally, since our results of Lemma~\ref{terrible} are valid for general intermediate  scores $(\alpha,\beta)$, an extension to the case where the initial score differs from $(0,0)$ is readily implemented.  This naturally paves the way to a more dynamic analysis of a match, \emph{\`a la} Klaassen and Magnus (2003),  in which the winning probabilities and duration  are sequentially estimated throughout the course of a given match. 

\begin{center}{\bf Acknowledgments } \end{center}
The authors  wish to thank the three anonymous referees for their helpful and insightful comments which have allowed for substantial improvement of the original manuscript.

\appendix

\section{Appendix.}

\subsection{Complete formulae.}
\begin{lemma}\label{horror}
Fix $m\in\Nn_0$ and $\al,\be\in\Nn$, and define $K,R\in\Nn$ via the Euclidian division $\al+\be=Km+R$. Then we have
\begin{enumerate}
\item if $R=0$ and $K$ is even, then
$$p_{\pla}^{\al,\be,\pla}(x)= \displaystyle{\binom{\frac{K}{2}m}{x}p_a^x(1-p_a)^{\frac{K}{2}m-x}\binom{\frac{K}{2}m-1}{\alpha-x-1}(1-p_b)^{\alpha-x}p_b^{\frac{K}{2}m-\alpha+x}},$$
where $x \in \{\max(0,\alpha-\frac{K}{2}m),\ldots,\min(\alpha-1, \frac{K}{2}m)\}$,
and 
$$p_{\pla}^{\al,\be,\plb}(x)= \displaystyle{\binom{\frac{K}{2}m}{x}p_a^x(1-p_a)^{\frac{K}{2}m-x}\binom{\frac{K}{2}m-1}{\alpha-x}(1-p_b)^{\alpha-x}p_b^{\frac{K}{2}m-\alpha+x}},$$
where $x \in \{\max(0,\alpha-\frac{K}{2}m+1),\ldots,\min(\alpha, \frac{K}{2}m)\}$.\\

\item if $R=0$ and $K$ is odd, then
$$p_{\pla}^{\al,\be,\pla}(x)=\displaystyle{\binom{\lceil{\frac{K}{2}}\rceil m-1}{x-1}p_a^x(1-p_a)^{\lceil{\frac{K}{2}}\rceil m-x}\binom{\lfloor{\frac{K}{2}}\rfloor m}{\alpha-x}(1-p_b)^{\alpha-x}p_b^{\lfloor{\frac{K}{2}}\rfloor m-\alpha+x}},$$
where $x \in \{\max(1,\alpha-\lfloor{\frac{K}{2}}\rfloor m),\ldots,\min(\alpha, \lceil{\frac{K}{2}}\rceil m)\}$, 
and
$$p_{\pla}^{\al,\be,\plb}(x)=\displaystyle{\binom{\lceil{\frac{K}{2}}\rceil m-1}{x}p_a^x(1-p_a)^{\lceil{\frac{K}{2}}\rceil m-x}\binom{\lfloor{\frac{K}{2}}\rfloor m}{\alpha-x}(1-p_b)^{\alpha-x}p_b^{\lfloor{\frac{K}{2}}\rfloor m-\alpha+x}},$$
where $x \in \{\max(0,\alpha-\lfloor{\frac{K}{2}}\rfloor m),\ldots,\min(\alpha, \lceil{\frac{K}{2}}\rceil m-1)\}$.\\

\item if $R>0$ and $K$ is even, then
$$p_{\pla}^{\al,\be,\pla}(x)=\displaystyle{\binom{\frac{K}{2}m+R-1}{x-1}p_a^x(1-p_a)^{\frac{K}{2}m+R-x}\binom{\frac{K}{2}m}{\alpha-x}(1-p_b)^{\alpha-x}p_b^{\frac{K}{2}m-\alpha+x}},$$
where $x \in \{\max(1,\alpha-\frac{K}{2}m),\ldots,\min(\alpha, \frac{K}{2}m+R)\}$,
and
$$p_{\pla}^{\al,\be,\plb}(x)= 
\displaystyle{\binom{\frac{K}{2}m+R-1}{x}p_a^x(1-p_a)^{\frac{K}{2}m+R-x}\binom{\frac{K}{2}m}{\alpha-x}(1-p_b)^{\alpha-x}p_b^{\frac{K}{2}m-\alpha+x}},$$
where $x \in \{\max(0,\alpha-\frac{K}{2}m),\ldots,\min(\alpha, \frac{K}{2}m+R-1)\}$.\\

\item if $R>0$ and $K$ is odd, then
$$p_{\pla}^{\al,\be,\pla}(x)=\displaystyle{\binom{\lceil{\frac{K}{2}}\rceil m}{x}p_a^x(1-p_a)^{\lceil\frac{K}{2}\rceil m-x}\binom{\lfloor{\frac{K}{2}}\rfloor m + R -1}{\alpha-x-1}(1-p_b)^{\alpha-x}p_b^{\lfloor{\frac{K}{2}}\rfloor m+R-\alpha+x}},$$
where $x \in \{\max(0,\alpha-\lfloor{\frac{K}{2}}\rfloor m-R),\ldots,\min(\alpha-1, \lceil{\frac{K}{2}}\rceil m)\}$,
and
$$p_{\pla}^{\al,\be,\plb}(x)=\displaystyle{\binom{\lceil{\frac{K}{2}}\rceil m}{x}p_a^x(1-p_a)^{\lceil\frac{K}{2}\rceil m-x}\binom{\lfloor{\frac{K}{2}}\rfloor m + R -1}{\alpha-x}(1-p_b)^{\alpha-x}p_b^{\lfloor{\frac{K}{2}}\rfloor m+R-\alpha+x}},$$
where $x \in \{\max(0,\alpha-\lfloor{\frac{K}{2}}\rfloor m-R+1),\ldots,\min(\alpha, \lceil{\frac{K}{2}}\rceil m)\}$. 
\end{enumerate}
\end{lemma}

\subsection{Proof of the complete formulae.}

Since all the formulae follow a very similar pattern, we only provide formal proofs for two of the eight expressions, namely for $p_\pla^{\alpha,\beta,\pla}(x)$ when $R=0$ and $K$ is even and for $p_\pla^{\alpha,\beta,\plb}(x)$ when $R>0$ and $K$ is odd.

Suppose $\alpha$ and $\beta$ are such that $\alpha+\beta=Km$ with $K$ even, and player~$\pla$ scores the last point. In this setting, the last point is served by player~$\plb$, hence only the outcome of $\frac{K}{2}m-1$ serves of player~$\plb$ are random, while all the $\frac{K}{2}m$ serves of player~$\pla$ are random. Thanks to the independence of the rallies, we may consider the serves of player~$\pla$ and player~$\plb$ separately. Moreover, as explained in Section~\ref{lemmasect}, each sequence of serves corresponds to a binomial distribution, with parameters $(\frac{K}{2}m,p_a)$ for player~$\pla$ serving and parameters $(\frac{K}{2}m-1,1-p_b)$ for player~$\plb$ serving, where we adopt each time the point of view of player~$\pla$. Since player~$\pla$ scores~$x$ points on his/her own serve, he/she necessarily wins $\alpha-x$ rallies  initiated by player~$\plb$. The last point being considered apart, we are left with $\alpha-x-1$ successes for the binomial distribution with parameters $(\frac{K}{2}m-1,1-p_b)$. Combining all these facts yields the announced formula, and it remains to establish the domain of the possible values of~$x$.

As player~$\plb$ serves exactly $\frac{K}{2}m$ times, it is clear that, whenever $\alpha>\frac{K}{2}m$, necessarily player~$\pla$ has to score at least $\alpha-\frac{K}{2}m$ points on his/her own serve. Otherwise, he/she may very well score only on player~$\plb$'s serve. These two observations readily yield the lower bound for~$x$. Regarding the upper bound, as player~$\pla$ scores at least one point not served by himself/herself, he/she cannot score more than $\alpha-1$ points on his/her own serve. This directly yields the upper bound, as player~$\pla$ is limited to $\frac{K}{2}m$ serves. 

Suppose now that $\alpha+\beta=Km+R$ with $K$ odd and $R>0$, and player~$\plb$ scores the last point. In this case, player~$\plb$ also serves the last point; would $R$ be zero, then the oddness of $K$ would give the last serve to player~$\pla$. With this said, player~$\pla$ serves exactly $\lceil\frac{K}{2}\rceil m$ points, while player~$\plb$ initiates $\lfloor\frac{K}{2}\rfloor m+R$ exchanges. The formula for $p_\pla^{\alpha,\beta,\plb}(x)$ and the corresponding lower and upper bound for~$x$ follow along the same lines as for $p_\pla^{\alpha,\beta,\pla}(x)$, which concludes the proof. 
\cqfd\\

\section*{References}

\noindent W.H. Carter $\&$ S.L. Crews (1974). An analysis of the game of tennis. \textit{The American Statistician}, 28, 130-134.

\noindent S.L. George (1973). Optimal strategy in tennis: a simple probabilistic model. \textit{Journal of the Royal Statistical Society Series C}, 22, 97-104.

\noindent   B.P. Hsi   $\&$ D.M.  Burich (1971). Games of two players.   \textit{Journal of the Royal Statistical Society Series C}, 20, 86-92.

\noindent F.J.G.M. Klaassen  $\&$  J.R. Magnus (2001). Are points in tennis independent and identically distributed? Evidence from a dynamic binary panel data model. \textit{Journal of the American Statistical Association}, 96, 500--509.

\noindent  F.J.G.M. Klaassen  $\&$  J.R. Magnus (2003).  Forecasting the winner of a tennis match.  \textit{European Journal of Operations Research}, 148, 257--267.

\noindent D. Paindaveine $\&$ Y. Swan (2011). A stochastic analysis of some two-person sports. \textit{Studies in Applied Mathematics, to appear}.

\noindent D.F. Percy (2009). A mathematical analysis of badminton scoring systems. \textit{Journal of the Operational Research Society}, 60, 63-71.

\noindent M.J. Phillips (1978). Sums of random variables having the modified geometric distribution with application to two-person games. \textit{Advances in Applied Probability}, 10, 647-665.

\noindent G.H. Riddle (1988). Probability Models for Tennis Scoring Systems. \textit{Journal of the Royal Statistical Society Series C}, 37, 63-75.

\noindent R.S. Schulman $\&$ M.A. Hamdan (1977). A probabilistic model for table tennis. \textit{The Canadian Journal of Statistics}, 5, 179-186.

\noindent J. Simmons (1989). A probabilistic model of squash: strategies and applications. \textit{Journal of the Royal Statistical Society Series C}, 38, 95-110.

\end{document}